\documentclass[aps,prl,twocolumn,superscriptadress,showpacs,amsmath,nobalancelastpage,letter,final]{revtex4}
\usepackage{lscape}
\usepackage{graphicx,latexsym}

\begin{document}
\begin{abstract}
We present numerical investigations of the dynamics on the energy landscape of
a realistic model of  the
high-temperature ceramic a-Si$_3$B$_3$N$_7$. Below a  critical
temperature $T_c \approx 2000$ K the system is no longer in equilibrium, and we 
predict that the material has a
glass transition  in this temperature range at high pressure. Analyzing the two-time
energy correlation function shows aging in this system, which is linked
to  the geometrical properties of the energy landscape.
\end{abstract}
\title{Non-equilibrium dynamics in amorphous Si$_3$B$_3$N$_7$ }
\author{A. Hannemann}
\affiliation{Max-Planck-Institut
f\"ur Festk\"orperforschung, Heisenbergstr. 1, D-70569 Stuttgart, Germany}
\author{J.C. Sch\"on}
\affiliation{Max-Planck-Institut
f\"ur Festk\"orperforschung, Heisenbergstr. 1, D-70569 Stuttgart, Germany}
\author{M. Jansen}
\affiliation{Max-Planck-Institut f\"ur Festk\"orperforschung, Heisenbergstr. 1, D-70569 Stuttgart, Germany}
\author{P. Sibani}
\affiliation{Fysisk Institut, SDU, Campusvej 55, DK-5230 Odense, Denmark}
\pacs{02.50.r, 05.90.+m, 61.41.+e, 61.43.Fs,
64.70.Pf}
\maketitle

\noindent {\bf Introduction: \ }
Amorphous nitridic ceramics containing both silicon and boron such as a-SiBN$_3$C 
or a-Si$_3$B$_3$N$_7$ are  synthesized via the sol-gel
process\cite{Jansen97a} and are one of the most exciting new classes
 of high-temperature materials. Up to now, no crystalline form of these compounds
is known. 
 Under standard conditions  a-SiBN$_3$C is thermally
stable and amorphous up to $\approx 2100$ K, possesses excellent
 mechanical and elastic properties, e.g.\ a
high bulk modulus of ca.\ $200 - 300$ GPa, and is stable against oxidation in
O$_2$-atmosphere up to $1700$ K\cite{Jansen97a}.\  
These compounds appear to form covalent networks with a homogeneous distribution of the
cations at least down to a scale of 1 nm\cite{Wullen00b}  
and  pose many fascinating questions 
regarding  their structure and physical properties.

Probing   the  glass  transition of a-Si$_3$B$_3$N$_7$,
the  basic representative of nitridic ceramics, 
is experimentally  difficult, since, under 
standard conditions, decomposition takes place  at $T\approx 1900$ K,
i.e.\  before the ceramic melts. This material is currently considered for 
high temperature engine applications and its aging properties are
therefore of clear technological relevance, beside having  
theoretical interest within the general framework of glassy dynamics.
 
Determining   whether a  possibly non-ergodic  system has
"for all practical  purposes" 
reached thermal (quasi-)equilibrium 
is not straightforward and possibly constitutes
an ill-posed question. Physical properties of
amorphous systems are  known to drift with the
time $t_w$, or age,  elapsed since the quench into the glassy 
phase. For short observation times
$t_{obs} \ll t_w $,  the drift is undetectable 
and a  state of quasi-equilibrium is revealed 
by the approximate validity of the
fluctuation-dissipation theorem. 
Concomitant  to the violations of the fluctuation-dissipation
theorem for $t_{obs} > t_w$,  the correlation and response functions
acquire an additional  dependence 
on  $t_w$.   This 
 breaking of time translational invariance
  has been  observed e.g. in the  magnetic susceptibility of spin
glasses, both in experiments\cite{Nordblad97a} and   model
simulations\cite{Hoffmann88a,Bouchaud97a},
in  measurements of $C_p$ for 
a-Se\cite{Stephens76a}, and also in simulations of the dynamical
structure factor of e.g.\ a-SiO$_2$ above the glass
transition temperature\cite{Kob99a}. 

To detect ergodicity breaking we  use 
1) the specific heat 
$C_V$, which we calculate in 
three  different ways, all agreeing  in equilibrium but 
markedly differing if ergodicity is broken, and  2) the
two-time energy-energy average $\phi(t_w, t_{obs};T)$,
 and the  related 
two-time autocorrelation function $C_E(t_w,t_{obs};T)$.
In quasi-equilibrium, the former  equals  one and the latter
equals  a generalized   standard 
equilibrium specific heat
$ k_BT^2C_V(t_w,t_{obs};T)$.
The age dependent $C_V$  has been studied experimentally, e.g.\ for charge-density-wave 
systems\cite{Biljakovic91a}, but does not appear to have been theoretically 
explored outside of two-level systems at very low temperatures\cite{Parshin93a}.

Since aging  is linked to the  complexity of the energy landscape of
the system, we have  investigated some aspects of the latter, emphasizing
their relation to the non-equilibrium dynamics.

\noindent {\bf Model and Techniques: \ }
The model of a-Si$_3$B$_3$N$_7$ consisted of $162$ Si-atoms, $162$ B-atoms and
$378$ N-atoms, respectively, in a $19.1 \times 19.1 \times 19.1$ {\AA}$^3$
cubic box. As an interaction potential, we employed a two-body potential from
the literature\cite{Gastreich00b} based on an-initio energy calculations
of hypothetical ternary compounds which reproduces  experimental data regarding 
the structure and vibrational properties of the binary compounds
Si$_3$N$_4$ and  BN and of molecules containing 
Si$-$N$-$B units. 

The starting configurations   for our  simulations were
generated by relaxation from  high temperature melts~\cite{Hannemann00a}.
The simulations were performed at  fixed temperature and volume, with a
Monte-Carlo algorithm using
the Metropolis acceptance criterion. 
In each update,  an atom is randomly selected for 
an attempted move in a random direction,  and with an average  size  
 chosen to achieve  an acceptance rate
of $\approx 50$ \%. One Monte-Carlo
cycle  (MCC) corresponds  to $N_{atom} =702$ such  individual
moves.
Note  that  
the kinetic energy ($3/2 k T$ per atom) does not 
appear in  MC-simulations, and that all   quantities studied 
relate to the configurational energy.

The temperatures investigated ranged from 25  to
7000 K. For each temperature up to $3000$ K and above $3000$ K, $9$ and $3$
runs, respectively, of length $t_{total} = 2 \times 10^5$  MCC  were performed.
 In addition, for selected temperatures, ensembles of
$100$ runs of length $t_{total} = 10^6$ MCC were studied. The 
energy as function of time  was
registered every $10$ MCC. Along the individual trajectories for $T =
250,...,7000$ K, halting points $x_H$ were chosen, from which both
conjugate gradient minimizations ($x_H \rightarrow x_{min}^{(1)}$) and a set of
$10$ stochastic quenches ($T=0$ K MC-runs) followed by conjugate gradient
minimizations ($x_H \rightarrow x_Q \rightarrow x_{min}^{(2)}$) were performed.
In the following,  $t_{init} \approx 1000$ MCC is the initialization
time of the MC-simulations needed for the system to reach equilibrium in the 
ergodic
regime (i.e.\ at high temperatures), while $t_w \ge t_{init}$ is the waiting 
time before the observation  begin.

\noindent {\bf Ergodicity : \ }
\label{Section_ergodicity}
To investigate   ergodicity versus  
temperature, we  studied   the specific heat $C_V$ and
the two-time energy-energy average
\begin{equation}
\phi({t_w},t_{obs};T) = \frac{\langle E(t_w)E(t_w + 
t_{obs})\rangle_{ens}}{\langle E(t_w)E(t_w)\rangle_{ens}}(T) \label{eq1},
\end{equation}
  where the subscript "ens" always  denotes an  average
over all trajectories. $C_V$ was calculated using three different computational
prescriptions.
First
\begin{equation}
C_V^a(T)    =   \frac{\partial \langle \langle E \rangle_{t 
\in[t_{w},t_{total}]}\rangle_{ens}(T)}{\partial T}, \label{eq2} 
\end{equation}
where  time averaging extends  from $t_w$ to 
the end of the simulation $t_{total}$ and the temperature
derivative is performed after the averaging.
Secondly 
\begin{eqnarray}
C_V^b (T) &=& \frac{\langle E(T+\Delta T;t_w) \rangle_{t \in [t_w,t_w + 
t_{obs}]; t_{obs} \ll t_w}}{2 \Delta T} \nonumber \\
&-&\frac{\langle E(T-\Delta T;t_w)\rangle_{t \in [t_w,t_w + t_{obs}];t_{obs} 
\ll t_w}}{2 \Delta T}.
\end{eqnarray}
This emulates a step experiment where  the system 
 ages  at temperature $T$. The
    time averages over  the observation time
    $t_{obs} \ll t_w$ are performed at temperatures
$T \pm \Delta T$,
where  $\Delta T \approx 0.1 T$.
Finally we gauge the energy fluctuations in
$[t_w,t_w + t_{obs}]$ by calculating  
 \begin{equation}
C_V^c (T) =  \frac{\langle \langle E^2\rangle_{t \in [t_w,t_w + t_{obs}]}         
- \langle E \rangle_{t \in [t_w,t_w + t_{obs}]}^2 \rangle_{ens} 
(T)}{k_BT^2}\label{eq4},
\end{equation}  
 for a range of observation times $t_{obs}$ which  straddles 
 $t_w$.

\begin{figure}
\includegraphics[width=0.53\columnwidth,angle=-90]{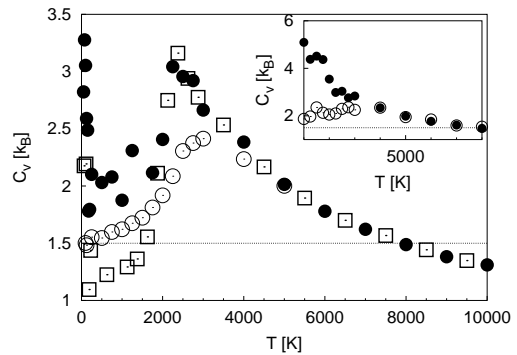}
\caption{Temperature dependence of the specific heats $C_V^{a,b,c}$. The
waiting $t_w$ and the observation times $t_{obs}$ were $10^5$ MCC for
$C_V^a$ ($\Box$)and $C_V^c$ $\bullet$, and $t_w \ge 10^5$, $t_{obs}=5\cdot10^3$ for $C_V^b$
($\circ$). Inset (note the different y-scale): $C_V^c$ for $t_w = 10^4$, $t_{obs}= 10^4$ ($\circ$),$10^5$ ($\bullet$).
Note that $C_V^c \approx C_V^b$ for $t_{obs} \leq t_w$, while for $t_{obs} > t_w$ the two
quantities differ.
\label{fig1}} 
\end{figure}

$C_V^a$  has  no  $t_{obs}$ dependence. By way of contrast, when  increasing 
$t_{obs}$ past $t_w$  the observed
dynamics in $C_V^c$
changes from quasi-equilibrium to off-equilibrium (cf. inset in fig.\ref{fig1}).
$C_V^b$, which mimicks  an
experiment performed after some relatively long equilibration time $t_w$, 
likely yields 
the most "realistic" value for the specific heat for all temperatures. 
\begin{figure}
\includegraphics[width=0.53\columnwidth,angle=-90]{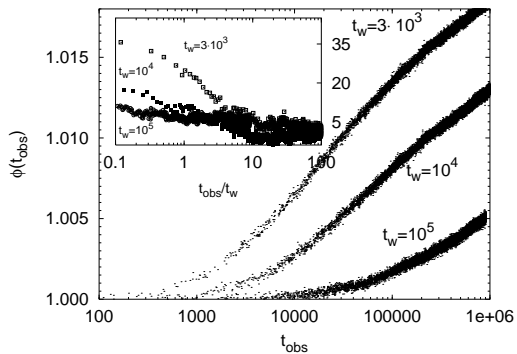}
\caption{Observation and waiting time dependences of the two-time energy-energy
average $\phi(t_w,t_{obs};T)$ for $1250$ K, for an ensemble size of
$100$ runs (raw data). The inset shows the two-time autocorrelation function
 $C_E(t_w,t_{obs};T=1250K)$.
Since, even for 100 runs (corresponding to ca. one year of CPU time on an AMD
1800+ MP processor), the scatter in $C_E$ is relatively large, the data in the inset are
averaged over ten time steps.
\label{fig2}} \end{figure}
As shown in fig.~\ref{fig1} the  above  prescriptions
yield, as expected,  almost identical results in the   high-$T$
ergodic dynamical regime, but differ at low  $T$. This indicates that 
 below $T_c \approx 2000 - 3000$ K ergodicity is broken. 
Further evidence  stems from the    observation that for  $T<T_c$, the 
motion is subdiffusive, while for $T>T_c$ standard diffusion is observed:
For $T > T_c\approx 2100$ K, the diffusion coefficients for  B, Si and N,
 follow a power law $D \propto(T-T_c)^{\gamma}$ with $\gamma = 1.7$, showing
   structural freezing-in.
Similarly, the relaxation times   associated with
 the bond survival probabilities  of B-N and Si-N
bonds display a rapid increase  below $T_c$.

Repeating these investigations for a large number of volumes,
we find a similar freezing-in of the structure for $T\approx T_c$.
Furthermore, we determine a critical point in the   liquid-gas region
of the ternary system 
($p_{cr} \approx 0.7$ GPa and, $T_{cr}  \approx 3700 $K ). 
Since the tendency to decompose is greatly reduced for a supercritical fluid,
we predict that a-Si$_3$B$_3$N$_7$ should exhibit a
glass transition at a  temperature $T_G\approx 1700 - 2000 $ K and at a pressure of
$p_G > 1$ GPa. 
Up to now, high-pressure experiments have only been performed for $T 
\approx 1000$ K and $p \approx 2.5$ GPa. 

For  $T>T_c$, the two-point correlation function 
always  remains very   close to
the equilibrium value  $1$. The aging behavior in the glassy phase 
is shown 
in fig.~\ref{fig2} for  $T=1250$ K, and for
three different waiting times $t_w =  3 \cdot 10^3, 10^4, 10^5$.
In the non-equilibrium regime 
$t_{obs} \ge t_w$,  $\phi$ is seen to  
deviate  strongly from its equilibrium value $\phi_{eq}\equiv 1$. 
The closely related autocorrelation function 
$C_E(t_w,t_{obs};T) \equiv \langle
E(t_w) \cdot E(t_w+t_{obs})\rangle_{ens}- \langle
E(t_w)\rangle_{ens}\cdot\langle E(t_w+t_{obs})\rangle_{ens}$ 
also exhibits the expected aging behavior, i.e. a
marked decrease to zero from an almost  constant value ($\propto C_V(t_w)$) once 
$t_{obs}$ exceeds $t_w$.
 This monotonic dependence on $t_w$ of the time range $t_{obs}
\in [0,t_w]$ during
which (quasi-)equilibrium behavior is still observed, correlates with the 
stiffening of
the response of the system characteristic for aging processes: The longer the 
system is
allowed to equilibrate, the longer is the subsequent time range  during which
equilibrium-like behavior is observed. This effect concurs with our 
observation
that for $T \le T_c$  we can fit $E(t;T)$ (and also $\langle 
E(t;T)\rangle_{ens}$) over the interval $[t_{init},t_{total}]$ as a
logarithmically decreasing function, $E(t;T) = E_0(T) - 
A(T)\ln\left(\frac{t}{t_0(T)}\right).$
Neglecting the fluctuations compared to the drift, 
one has 
$\phi(t_w,t_{obs};T) \approx  \frac{E(t_w+t_{obs})E(t_w)}{E(t_w)E(t_w)}$.
Expanding $\phi$ for $t_{obs} \ll t_w$ then yields  
$\phi(t_w.t_{obs};T) \approx 1 + 
\frac{A}{|E(t_w)|}\frac{t_{obs}}{t_w}$.
Thus, $\phi(t_w,t_{obs};T)$   
 substantially deviates from $1$  for  $t_{obs} >  t_w$, as
observed in the simulations. The inset shows $C_E(t_w,t_{obs};T)$ plotted as function of
the scaled variable $t_{obs}/t_w$. As $t_{obs}$ increases, the data appears to collapse
on a single curve, indicating that $t_{obs}/t_w$ scaling can be expected to hold
asymptotically.

\begin{figure}
\includegraphics[width=0.53\columnwidth,angle=-90]{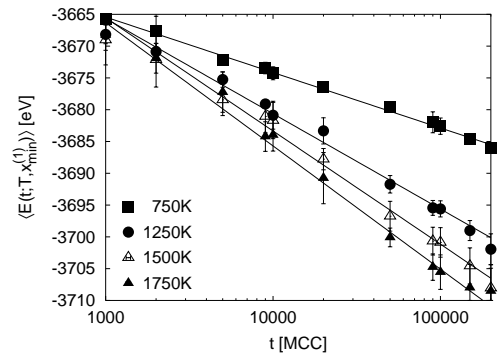}  
\caption{Time dependence of the average energies
$\langle E(t;T,x_{min}^{(1)})\rangle_{ens}$ of the minima $x_{min}^{(1)}$ for selected
temperatures $T=750$ K, $1250$ K, $1500$ K, $1750$ K. \label{fig3}} 
\end{figure}

\noindent {\bf Energy landscape: \ } 
Finally, we would like to link the  non-equilibrium behavior 
to the properties of the energy landscape of a-Si$_3$B$_3$N$_7$.
Figure \ref{fig3} shows the average energy
$\langle E(t;T,x_{min}^{(1)})\rangle_{ens}$ of local
minima  $x_{min}^{(1)}$ found by applying
a  conjugate gradient   algorithm  
 for logarithmically
spaced halting points along several trajectories as a function of time for
different temperatures. 

We note that $\langle E(t;T,x_{min}^{(1)})\rangle_{ens}$
decreases logarithmically with time for $T < T_c$ analogously
to $\langle E(t;T) \rangle_{ens}$~\cite{Angelani01a}. 
A fit of the logarithmic slope yields $A(T) = 76.29 \cdot T -  134.56\cdot
T^2$, which   qualitatively agrees 
with the   low temperature expansion of $\langle E(t;T)
\rangle_{ens}$  for the so-called LS-tree
models\cite{Sibani91a}, suggesting  that the landscape of a-Si$_3$B$_3$N$_7$
might possess some hierarchical aspects in
that energy range relevant for   $T <T_c$.

\begin{figure}
\includegraphics[width=0.53\columnwidth,angle=-90]{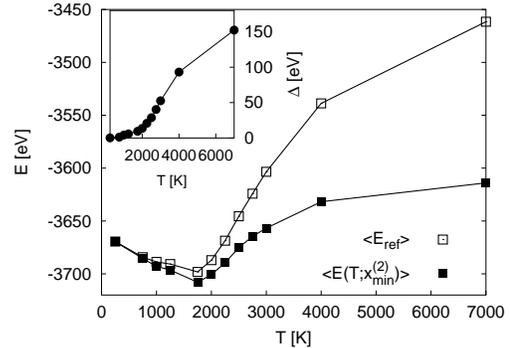}
\caption{Temperature dependence of the average energies $\langle \langle E(T;x_{min}^{(2)})\rangle_{t_{obs}}\rangle_{ens}$
of the
minima $x_{min}^{(2)}$ (Full squares). The open squares are values of reference
energies $E_{ref}(T)$ (see text). The inset shows the temperature dependence
of the excess energy $\Delta$ (see text).\label{fig4}}
\end{figure}

For fixed simulation time, the deepest local minima are reached for $T = 1750$
K, which
lies right below $T_c$. We find a similar behavior for the average energy $\langle E(T;x_{min}^{(2)})\rangle$
of the local minima $x_{min}^{(2)}$ found after quenching plus gradient minimization starting from the
holding points $x_h$, shown as a function of temperature in fig.\ \ref{fig4}. We clearly recognize a
minimum in this curve at $T \approx 1750$ K, and the largest increase occurs at $T \approx T_c$.
Analogous observations are well-known from e.g.\ global optimization studies of complex systems, where it
has been found that reaching the deepest local minima using Monte-Carlo-type search algorithms is
achieved by spending most of the search time in the temperature interval slightly below the glass
transition temperature\cite{Kirkpatrick83a}. Thus this result serves as another confirmation that ergodicity
breaking is taking place at $T \approx T_c$.


The second curve in fig.\ \ref{fig4} depicts
$E_{ref}(T) = \langle \langle E(T,x_h)\rangle_{t_{obs}}\rangle_{ens} - (3/2)N_{atom}k_BT. $
 Note that if the motion around the minima were purely
 harmonic, $E_{ref}(T)$ would be the average energy of these minima.  
However, comparing $E_{ref}(T)$ with $\langle \langle E(T,x_{min}^{(2)})\rangle_{t_{obs}}\rangle_{ens}$,
 we note that an excess energy
$\Delta(T) = E_{ref}(T) - \langle \langle E(T,x_{min}^{(2)})\rangle_{t_{obs}}\rangle_{ens} > 0$
exists. The  growth of $\Delta(T)$ is  monotonic in $T$  and most pronounced for 
$T \approx T_c$. Clearly, in this temperature region the landscape below the halting points, which
is "felt" by the walker, changes and a substantial shift to higher "reference" values for the
vibrational contributions to the potential energy occurs.

\noindent {\bf Discussion: \ }
The computational analysis  of a-Si$_3$B$_3$N$_7$ using $C_V^{a,b,c}$ and $\phi(t_w, t_{obs};T)$
 shows that this amorphous material can be expected to 
 exhibit a glass transition with a concurrent break in 
 the ergodicity at about $T_c \approx 2000$ K if
 pressures high enough to prevent decomposition are applied. 
 Regarding its structural dynamics and aging properties 
 for $T < T_c$, a-Si$_3$B$_3$N$_7$ exhibits a  general behavior 
 similar to standard test systems (Lennard-Jones, a-SiO$_2$) and CDW
 systems,  
 insofar as we observe a freezing-in of the structure, 
 and a waiting-time dependence of the two time correlation function 
 and specific heat. 

This aging phenomenon is related to the slow non-exponential relaxation dynamics
 on the energy landscape for $T < T_c$, resulting in a logarithmic drift
  towards lower energies. This applies both to the actual trajectories 
  and the time-sequence of observed local minima. 
  Independent of this aspect of the dynamics, we find that starting around 
  $T_c$ the average potential energy greatly exceeds the value associated
   with harmonic vibrations at $T_c$. 
   'Thermodynamically', this makes itself felt as a peak in the specific heat, 
   which is often associated with the entropy due to 
   an increased availability of additional amorphous configurations.

One possible origin for this excess energy is 
  trapping~\cite{Sibani93a,Schoen97b} due to an
approximately exponential growth in the effective local density of 
states $g_{loc}(E) \propto\exp(\alpha (E - E_{min}))$, with the growth factor $\alpha = 1/T_{trap} \approx 1/T_c$. In such a
case, the exponentially growing region of the deep pockets of the landscape becomes invisible for
the random walker for $T > T_{trap}$, and the top of this region serves as the reference for the
vibrational energy contribution. $\Delta (T)$ would thus refer to the depth of the
exponentially growing pockets with $T_{trap} < T$. Studies of 
amorphous networks\cite{Schoen00a} and polymers\cite{Schoen02a}
 on lattices have suggested that such trapping  might contribute to
the glass transition in structural glasses.

Alternatively, the phase space volume of the locally ergodic regions
 around the saddle points might substantially exceed the one associated
  with local minima, such that these saddles serve as
reference, as has been suggested for small Lennard-Jones systems\cite{Cavagna01a}.
 Here, $\Delta (T)$ would correspond to the average energy difference
between the relevant saddle points and the nearby local minima. 

Landscape studies using the threshold
algorithm\cite{Schoen96a},  which can be used to estimate both local densities of states
and minima, and the relative
size of minimum and transition regions\cite{Wevers00a} could resolve this issue, but are not yet computationally
feasible. Preliminary explorations of 
the region "below" the halting points by performing 10 quench runs
for each halting point $x_h^{(i)}$ concur with the results obtained in\cite{Wevers00a} and
show that a typical halting point is associated with only one
rather circumscribed basin containing many similar minima, and not with a large region of the
landscape encompassing very different structures. The latter would be expected if the dynamics
were dominated by high-lying saddle points.
\begin{acknowledgments}
Funding was kindly provided by the DFG
via SFB408.
\end{acknowledgments}
\bibliography{g42}
\end{document}